\documentstyle[12pt,graphicx]{article}

\def\mathfrak{\bf}

\def\be{\begin{equation}}
\def\ee{\end{equation}}
\def\bea{\begin{eqnarray}}
\def\eea{\end{eqnarray}}

\def\dt#1{\on{\hbox{\bf .}}{#1}}                
\def\Dot#1{\dt{#1}}
\def\IR{\relax{\rm I\kern-.18em R}}
\def\binomial#1#2{\left(\,{\buildrel 
{\raise4pt\hbox{$\displaystyle{#1}$}}\over
{\raise-6pt\hbox{$\displaystyle{#2}$}}}\,\right)}

\def\[{\lfloor{\hskip 0.35pt}\!\!\!\lceil}
\def\]{\rfloor{\hskip 0.35pt}\!\!\!\rceil}

\newcommand{\AmS}{{\protect\the\textfont2
  A\kern-.1667em\lower.5ex\hbox{M}\kern-.125emS}}

\catcode`@=11
\def\un#1{\relax\ifmmode\@@underline#1\else
        $\@@underline{\hbox{#1}}$\relax\fi}
\catcode`@=12

\def\ad{{\kern0.5pt
                   \alpha \kern-5.05pt
\raise5.8pt\hbox{$\textstyle.$}\kern
0.5pt}}

\def\Dot#1{{\kern0.5pt
     {#1} \kern-5.05pt \raise5.8pt\hbox{$\textstyle.$}\kern
0.5pt}}

\def\bo{{\raise.15ex\hbox{\large$\Box$}}}               
\def\TH{{\raise.2ex\hbox{$\displaystyle \bigodot$}\mskip-4.7mu \llap H
\;}}
\def\face{{\raise.2ex\hbox{$\displaystyle \bigodot$}\mskip-2.2mu \llap
{$\ddot
        \smile$}}}                                      

\def\Tilde#1{\widetilde{#1}}                    
\def\leftrightarrowfill{$\mathsurround=0pt \mathord\leftarrow \mkern-6mu
        \cleaders\hbox{$\mkern-2mu \mathord- \mkern-2mu$}\hfill
        \mkern-6mu \mathord\rightarrow$}
\def\dvec#1{\vbox{\ialign{##\crcr
        \leftrightarrowfill\crcr\noalign{\kern-1pt\nointerlineskip}
        $\hfil\displaystyle{#1}\hfil$\crcr}}}           
\def\dt#1{{\buildrel {\hbox{\LARGE .}} \over {#1}}}     

\def\frac#1#2{{\textstyle{#1\over\vphantom2\smash{\raise.20ex
        \hbox{$\scriptstyle{#2}$}}}}}                   
\def\sfrac#1#2{{\vphantom1\smash{\lower.5ex\hbox{\small$#1$}}\over
        \vphantom1\smash{\raise.4ex\hbox{\small$#2$}}}} 
\def\bfrac#1#2{{\vphantom1\smash{\lower.5ex\hbox{$#1$}}\over
        \vphantom1\smash{\raise.3ex\hbox{$#2$}}}}       
\def\afrac#1#2{{\vphantom1\smash{\lower.5ex\hbox{$#1$}}\over#2}}    
\def\on#1#2{\mathop{\null#2}\limits^{#1}}               

\newskip\humongous \humongous=0pt plus 1000pt minus 1000pt
\def\caja{\mathsurround=0pt}
\def\eqalign#1{\,\vcenter{\openup2\jot \caja
        \ialign{\strut \hfil$\displaystyle{##}$&$
        \displaystyle{{}##}$\hfil\crcr#1\crcr}}\,}
\newif\ifdtup

  \def\pp{{\mathchoice
          {
              \kern 1pt%
              \raise 1pt
              \vbox{\hrule width5pt height0.4pt depth0pt
                    \kern -2pt
                    \hbox{\kern 2.3pt
                          \vrule width0.4pt height6pt depth0pt
                          }
                    \kern -2pt
                    \hrule width5pt height0.4pt depth0pt}%
                    \kern 1pt
           }
            {
              \kern 1pt%
              \raise 1pt
              \vbox{\hrule width4.3pt height0.4pt depth0pt
                    \kern -1.8pt
                    \hbox{\kern 1.95pt
                          \vrule width0.4pt height5.4pt depth0pt
                          }
                    \kern -1.8pt
                    \hrule width4.3pt height0.4pt depth0pt}%
                    \kern 1pt
            }
            {
              \kern 0.5pt%
              \raise 1pt
              \vbox{\hrule width4.0pt height0.3pt depth0pt
                    \kern -1.9pt  
                    \hbox{\kern 1.85pt
                          \vrule width0.3pt height5.7pt depth0pt
                          }
                    \kern -1.9pt
                    \hrule width4.0pt height0.3pt depth0pt}%
                    \kern 0.5pt
            }
            {
              \kern 0.5pt%
              \raise 1pt
              \vbox{\hrule width3.6pt height0.3pt depth0pt
                    \kern -1.5pt
                    \hbox{\kern 1.65pt
                          \vrule width0.3pt height4.5pt depth0pt
                          }
                    \kern -1.5pt
                    \hrule width3.6pt height0.3pt depth0pt}%
                    \kern 0.5pt
            }
        }}

\def\pd{{\kern0.5pt
                   + \kern-5.05pt \raise5.8pt\hbox{$\textstyle.$}\kern
0.5pt}}

\def\pmd{{\kern0.5pt
                  \pm \kern-5.05pt \raise6.3pt\hbox{$\textstyle.$}\kern1.5pt}}

\def\md{{\mathchoice
   {
      {{\kern 1pt - \kern-6.2pt \raise5pt\hbox{$\textstyle.$}\kern 1pt}}}
    {
      {{\kern 1pt - \kern-6.2pt \raise5pt\hbox{$\textstyle.$}\kern 1pt}}}
    {
      {\kern0.5pt - \kern-5.05pt \raise3.4pt\hbox{$\textstyle.$}\kern0.5pt}}
    {
      {\kern0.5pt - \kern-5.05pt \raise3.4pt\hbox{$\textstyle.$}\kern0.5pt}}}}

\def\ad{{\dot{\alpha}}}

\def\pp{{\mathchoice
          {
              \kern 1pt%
              \raise 1pt
              \vbox{\hrule width5pt height0.4pt depth0pt
                    \kern -2pt
                    \hbox{\kern 2.3pt
                          \vrule width0.4pt height6pt depth0pt
                          }
                    \kern -2pt
                    \hrule width5pt height0.4pt depth0pt}%
                    \kern 1pt
           }
            {
              \kern 1pt%
              \raise 1pt
              \vbox{\hrule width4.3pt height0.4pt depth0pt
                    \kern -1.8pt
                    \hbox{\kern 1.95pt
                          \vrule width0.4pt height5.4pt depth0pt
                          }
                    \kern -1.8pt
                    \hrule width4.3pt height0.4pt depth0pt}%
                    \kern 1pt
            }
            {
              \kern 0.5pt%
              \raise 1pt
              \vbox{\hrule width4.0pt height0.3pt depth0pt
                    \kern -1.9pt  
                    \hbox{\kern 1.85pt
                          \vrule width0.3pt height5.7pt depth0pt
                          }
                    \kern -1.9pt
                    \hrule width4.0pt height0.3pt depth0pt}%
                    \kern 0.5pt
            }
            {
              \kern 0.5pt%
              \raise 1pt
              \vbox{\hrule width3.6pt height0.3pt depth0pt
                    \kern -1.5pt
                    \hbox{\kern 1.65pt
                          \vrule width0.3pt height4.5pt depth0pt
                          }
                    \kern -1.5pt
                    \hrule width3.6pt height0.3pt depth0pt}%
                    \kern 0.5pt
            }
        }}

\def\pd{{\kern0.5pt
                   + \kern-5.05pt \raise5.8pt\hbox{$\textstyle.$}\kern
0.5pt}}

\def\pmd{{\kern0.5pt
                  \pm \kern-5.05pt \raise6.3pt\hbox{$\textstyle.$}\kern1.5pt}}

\def\md{{\mathchoice
   {
      {{\kern 1pt - \kern-6.2pt \raise5pt\hbox{$\textstyle.$}\kern 1pt}}}
    {
      {{\kern 1pt - \kern-6.2pt \raise5pt\hbox{$\textstyle.$}\kern 1pt}}}
    {
      {\kern0.5pt - \kern-5.05pt \raise3.4pt\hbox{$\textstyle.$}\kern0.5pt}}
    {
      {\kern0.5pt - \kern-5.05pt \raise3.4pt\hbox{$\textstyle.$}\kern0.5pt}}}}

\def\dslash{\not{\hbox{\kern-2pt $\partial$}}}
\def\Dslash{\not{\hbox{\kern-4pt $D$}}}
\def\pslash{\not{\hbox{\kern-2.3pt $p$}}}
 \newtoks\slashfraction
 \slashfraction={.13}
 \def\slash#1{\setbox0\hbox{$ #1 $}
 \setbox0\hbox to \the\slashfraction\wd0{\hss \box0}/\box0 }
 
\font\ro=cmsy10                          
\def\kcr{{\hbox{\ro \char'170}}}                
\def\ktl{{\hbox{\ro \char'170}}}        
\def\ktr{{\hbox{\ro \char'170}}}        
\def\kbl{{\hbox{\ro \char'170}}}        
\def\kbr{{\hbox{\ro \char'170}}}        

\def\plpl{\raise-2pt\hbox{$\raise3pt\hbox{$_+$}\hskip-6.67pt\raise0.0pt
\hbox{$^+$}\hskip 0.01pt$}}
\def\mimi{\raise-2pt\hbox{$\raise3pt\hbox{$_-$}\hskip-6.67pt\raise0.0pt
\hbox{$^-$}\hskip 0.01pt$}} 

\def\bo{{\raise.15ex\hbox{\large$\Box$}}}               
\def\TH{{\raise.2ex\hbox{$\displaystyle \bigodot$}\mskip-4.7mu \llap H \;}}
\def\face{{\raise.2ex\hbox{$\displaystyle \bigodot$}\mskip-2.2mu \llap {$\ddot
        \smile$}}}                                      

\def\Tilde#1{\widetilde{#1}}                    
\def\leftrightarrowfill{$\mathsurround=0pt \mathord\leftarrow \mkern-6mu
        \cleaders\hbox{$\mkern-2mu \mathord- \mkern-2mu$}\hfill
        \mkern-6mu \mathord\rightarrow$}
\def\dvec#1{\vbox{\ialign{##\crcr
        \leftrightarrowfill\crcr\noalign{\kern-1pt\nointerlineskip}
        $\hfil\displaystyle{#1}\hfil$\crcr}}}           
\def\dt#1{{\buildrel {\hbox{\LARGE .}} \over {#1}}}     

\def\frac#1#2{{\textstyle{#1\over\vphantom2\smash{\raise.20ex
        \hbox{$\scriptstyle{#2}$}}}}}                   
\def\sfrac#1#2{{\vphantom1\smash{\lower.5ex\hbox{\small$#1$}}\over
        \vphantom1\smash{\raise.4ex\hbox{\small$#2$}}}} 
\def\bfrac#1#2{{\vphantom1\smash{\lower.5ex\hbox{$#1$}}\over
        \vphantom1\smash{\raise.3ex\hbox{$#2$}}}}       
\def\afrac#1#2{{\vphantom1\smash{\lower.5ex\hbox{$#1$}}\over#2}}    
\def\on#1#2{\mathop{\null#2}\limits^{#1}}               

\parskip=\medskipamount                 
\thicklines                         

\def\oldheadpic{                                
        \setlength{\unitlength}{.4mm}
        \thinlines
        \par
        \begin{picture}(349,16)
        \put(325,16){\line(1,0){4}}
        \put(330,16){\line(1,0){4}}
        \put(340,16){\line(1,0){4}}
        \put(335,0){\line(1,0){4}}
        \put(340,0){\line(1,0){4}}
        \put(345,0){\line(1,0){4}}
        \put(329,0){\line(0,1){16}}
        \put(330,0){\line(0,1){16}}
        \put(339,0){\line(0,1){16}}
        \put(340,0){\line(0,1){16}}
        \put(344,0){\line(0,1){16}}
        \put(345,0){\line(0,1){16}}
        \put(329,16){\oval(8,32)[bl]}
        \put(330,16){\oval(8,32)[br]}
        \put(339,0){\oval(8,32)[tl]}
        \put(345,0){\oval(8,32)[tr]}
        \end{picture}
        \par
        \thicklines
        \vskip.2in}
\def\oldtitle#1#2#3#4{\oldheadpic\begin{center}\vglue.5in{\large\bf #1}\\[.6in]
        {#2}\\[.1in] {\it Department of Physics and Astronomy}\\
        {\it University of Maryland, College Park, MD 20742}\\[.6in]
        Physics Publication \#{#3}\\ {#4}\\[1.5in] {\bf ABSTRACT}\\[.1in]
        \end{center} \begin{quotation}}                 
\def\oldTitle#1#2#3#4#5#6#7{\oldheadpic\begin{center} \vglue .4in
        {\large\bf #1}\\[.4in]
        {#2}\\[.1in] {\it Department of Physics and Astronomy}\\
        {\it University of Maryland, College Park, MD 20742}\\[.1in]
        {#3}\\[.1in] {\it {#4}}\\ {\it {#5}}\\[.4in]
        Physics Publication \#{#6}\\ {#7}\\[.5in] {\bf ABSTRACT}\\[.1in]
        \end{center} \begin{quotation}}                 
\def\border{                                            
        \setlength{\unitlength}{1mm}
        \newcount\xco
        \newcount\yco
        \xco=-21
        \yco=12
        \begin{picture}(140,0)
        \put(\xco,\yco){$\ktl$}
        \advance\yco by-1
        {\loop
        \put(\xco,\yco){$\kcr$}
        \advance\yco by-2
        \ifnum\yco>-240
        \repeat
        \put(\xco,\yco){$\kbl$}}
        \xco=158
        \yco=12
        \put(\xco,\yco){$\ktr$}
        \advance\yco by-1
        {\loop
        \put(\xco,\yco){$\kcr$}
        \advance\yco by-2
        \ifnum\yco>-240
        \repeat
        \put(\xco,\yco){$\kbr$}}
        \put(-20,13){\tiny *University of Maryland ** Center for String and 
         Particle  Theory ** Physics Department *** Stanford Linear Accelerator Center **** 
         Stanford University *}
        \put(-20,-241.5){\tiny *University of Maryland ** Center for String and 
         Particle  Theory ** Physics Department *** Stanford Linear Accelerator Center **** 
         Stanford University *}
        \end{picture}
        \par\vskip-8mm}
\def\bordero{                                           
        \setlength{\unitlength}{1mm}
        \newcount\xco
        \newcount\yco
        \xco=-31
        \yco=12
        \begin{picture}(140,0)
        \put(\xco,\yco){$\ktl$}
        \advance\yco by-1
        {\loop
        \put(\xco,\yco){$\kclr}
        \advance\yco by-2
        \ifnum\yco>-240
        \repeat
        \put(\xco,\yco){$\kbl$}}
        \xco=151
        \yco=12
        \put(\xco,\yco){$\ktr$}
        \advance\yco by-1
        {\loop
        \put(\xco,\yco){$\kcr$}
        \advance\yco by-2
        \ifnum\yco>-240
        \repeat
        \put(\xco,\yco){$\kbr$}}
        \put(-20,12){\ooo bacdefghidfghghdhededbihdgdfdfhhdheidhdhebaaahjhhdahba

hgdedge
   hgfdiehhgdigicba}
        \put(-20,-241.5){\ooo ababaighefdbfghgeahgdfgafagihdidihiidhiagfedhadbfd

ecdcdfa
   gdcbhaddhbgfchbgfdacfediacbabab}
        \end{picture}
        \par\vskip-8mm}
\def\headpic{                                           
        \indent
        \setlength{\unitlength}{.4mm}
        \thinlines
        \par
        \begin{picture}(29,16)
        \put(165,16){\line(1,0){4}}
        \put(170,16){\line(1,0){4}}
        \put(180,16){\line(1,0){4}}
        \put(175,0){\line(1,0){4}}
        \put(180,0){\line(1,0){4}}
        \put(185,0){\line(1,0){4}}
        \put(169,0){\line(0,1){16}}
        \put(170,0){\line(0,1){16}}
        \put(179,0){\line(0,1){16}}
        \put(180,0){\line(0,1){16}}
        \put(184,0){\line(0,1){16}}
        \put(185,0){\line(0,1){16}}
        \put(169,16){\oval(8,32)[bl]}
        \put(170,16){\oval(8,32)[br]}
        \put(179,0){\oval(8,32)[tl]}
        \put(185,0){\oval(8,32)[tr]}
        \end{picture}
        \par\vskip-6.5mm
        \thicklines}
\def\title#1#2#3#4{\border\headpic {\hbox to\hsize{#4 \hfill UMDEPP #3}}\par
        \begin{center} \vglue .5in {\large\bf #1}\\[.6in]
        {#2}\\[.1in] {\it Department of Physics and Astronomy}\\
        {\it University of Maryland, College Park, MD 20742}\\[1.5in]
        {\bf ABSTRACT}\\[.1in] \end{center} \begin{quotation}}  
\def\Title#1#2#3#4#5#6#7{\border\headpic
        {\hbox to\hsize{#7 \hfill UMDEPP #6}}\par
        \begin{center} \vglue .4in {\large\bf #1}\\[.4in]
        {#2}\\[.1in] {\it Department of Physics and Astronomy}\\
        {\it University of Maryland, College Park, MD 20742}\\[.1in]
        {#3}\\[.1in] {\it {#4}}\\ {\it {#5}}\\[.5in] {\bf ABSTRACT}\\[.1in]
        \end{center} \begin{quotation}}                 
\def\endtitle{\end{quotation}\newpage}                  

\def\qd{{\kern0.5pt
                   q \kern-5.05pt \raise5.8pt\hbox{$\textstyle.$}\kern
0.5pt}}

\newcommand{\bse}{\begin{subequations}}
\newcommand{\ese}{\end{subequations}}
\newcommand{\half}{{1\over2}}
\newcommand{\ba}{\begin{array}}
\newcommand{\ea}{\end{array}}
\newcommand{\eee}{\end{equation}}

\begin{document}

\def\dt#1{\on{\hbox{\bf .}}{#1}}                
\def\Dot#1{\dt{#1}}

\def\gfrac#1#2{\frac {\scriptstyle{#1}}
ÊÊÊÊÊÊÊ {\mbox{\raisebox{-.6ex}{$\scriptstyle{#2}$}}}}
\border\headpic {\hbox to\hsize{August 2004 \hfill
{SLAC-PUB-xxxxx}}}  
\par
{$~$ \hfill
{UMDEPP 05-014}}
\par
{$~$ \hfill
{hep-th/yymmnn}}
\par

\setlength{\oddsidemargin}{0.3in}
\setlength{\evensidemargin}{-0.3in}
\begin{center}
\vglue .10in
{\large\bf Can the String Scale be Related to the \\
Cosmic Baryon Asymmetry?\footnote
{Supported in part  by National Science Foundation Grant
PHY-0354401.}\  }
\\[.3in]

Stephon H. S, Alexander\footnote{stephon@itp.stanford.edu}
\\[0.05in]
{\it Stanford Linear Accelerator Center\\
Stanford University\\
2575 Sand Hill Road, Menlo Park CA,  USA}\\[.1in]
and 
\\[.1in]

S.\, James Gates, Jr.\footnote{gatess@wam.umd.edu}${}$
\\[0.05in]
{\it Center for String and Particle Theory\\ 
Department of Physics, University of Maryland\\ 
College Park, MD 20742-4111 USA}\\[.3in]

{\bf ABSTRACT}\\[.01in]
\end{center}
\begin{quotation}
{In a previous work, a mechanism was presented by which Baryon asymmetry 
can be generated during inflation from elliptically polarized gravitons.  
Nonetheless, the mechanism only generated a realistic baryon asymmetry 
under special circumstances which requires an enhancement of the lepton 
number from an unspecified GUT.   In this note we provide a stringy embedding 
of this mechanism demonstrating that if the model-independent axion is the source 
of the gravitational waves responsible for the lepton asymmetry, one can 
observationally constrain the string scale and coupling. 
} 

${~~~}$ \newline

\endtitle

\section{Introduction}

~~~~ Inflation stands as the paradigm within which the problems of the standard 
big bang cosmology are resolved.  Moreover, inflation provides a causal mechanism 
necessary for the formation of large scale structure (LSS) in the universe.  Recently 
one of the present authors (SHSA) in a collaboration \cite{peskin} presented a model
in which the quantum fluctuations of gravity waves generated during the inflationary epoch can  give rise to the cosmic matter-antimatter asymmetry.   This mechanism relies on a rolling pseudo-scalar field 
coupled to a gravitational Chern-Simons term.   However, a concrete model in 
the context of a fundamental theory was not specified in this work \cite{peskin}.  In 
the present note we provide an embedding of this leptogenesis mechanism into 
a 4D, ${\cal N} \,= \,1$ SUGRA limit of closed type-I superstring or heterotic string 
theory, in which the model independent axion is dynamical.   Since the axion is a 
universal field in any 4D, ${\cal N} \,= \,1$ string theory compactification, we shall 
obtain a unique constraint on the string scale from a cosmological observable, the 
baryon asymmetry index.

\subsection{Review of Inflationary Leptogenesis}

~~~~ The key to inflationary leptogenesis is the generation of a bi-refringent gravitational 
wave (BG)¥ spectrum during the course of inflation.   These BG waves are quantum 
mechanically produced during inflation if there is a non-vanishing axion correlated 
on the scale of the horizon at the beginning of inflation.  Subsequently, through a 
triangle ABJ anomaly, a non vanishing quantum expectation value for ${\cal R}\wedge 
{\cal R}$ is generated from the contraction of BG waves which were sourced by the 
non-vanishing axion.  When $<{{\cal R}\wedge {\cal R}}>$ integrated during the whole 
course of inflation, one gets a non-vanishing lepton number.  

It is well known from the ABJ anomaly that CP violating interactions at one loop can 
induce an gravitational Chern-Simons term.  The authors \cite{peskin} proposed a 
model where this term can arise from a coupling of a pseudo-scalar to the gravitational 
Chern-Simons term.  Our goal in this section is to briefly review this proposition by 
assuming such a term exists and finding explicit solutions of gravitational waves by 
linearizing the Einstein-Hilbert action.  We will then proceed to review the APS \cite{peskin} computation of the lepton 
number produced during inflation.

The authors \cite{peskin} considered a general coupling between the the axion and the CP violating curvature invariant.  
 
\be \label{axion-graviton}
\cal{L}\rm_{int} ~=~ F(\phi)\, {\cal R}^{\alpha}_{\ \sigma \mu \nu}\,  \Tilde{\cal 
R}_{\alpha}^{\ \sigma \mu 
\nu}. \ee
In terms of linearized perturbations up to second order this term is exactly\footnote{In \ref{lepton} the gravity waves are chosen to move in the z-direction with any loss of generality}.
\be\label{lepton} 
\eqalign{
{\rm Tr}({\cal R} \wedge {\cal R}) &=~ \frac{8i}{a^{3}}\,  \biggl[ ~
\left(\frac{\partial^2~}{\partial z^2}h_R \right) \left(
\frac{\partial^2 ~\, }{\partial t\partial z}h_L \right)  \,-\, 
\left( \frac{\partial^2~}{\partial z^2}h_L \right) \left(
\frac{\partial^2~\,}{\partial t\partial z}h_R\right) {~~~~~~~~~} \cr
&~~~~~~~~~~~~+~ a^2  \, {\Big\{} \left(\frac{\partial^2~}{\partial t^2}
h_R \right) \left(\frac{\partial^2~\,}{\partial t\partial z}h_L  \right)  \,-\, 
\left( \frac{\partial^2~}{\partial t^2}h_L \right) \left(
\frac{\partial^2~\,}{\partial t\partial z}h_R\right) \, {\Big\} } \cr
&~~~~~~~~~~~~+~ \left( \frac{1}{2}\frac{\partial~}{\partial t}a^2  \right) {\Big\{} \left( 
\frac{\partial ~}{\partial t}h_R  \right) \left(\frac{\partial^2~\,}{\partial t\partial z}
h_L   \right)  \,-\, \left(  \frac{\partial~}{\partial t}h_L  \right) \left(
\frac{\partial^2~\,}{\partial t \partial z}h_R\right) \, {\Big\} }~ \biggr]  
~~~, }
\ee 
where $h_{L}$ and $h_{R}$ are left and right helicities of the gravity waves respectively.
The resulting gravity waves will be sourced by the axion via. the equations of motion
\be\label{hphc-e.o.m}
\eqalign{
\Box h_L=& -i \,2 \frac{\Theta}{a}\, \dot h'_L \cr
\Box h_R=& +i \, 2 \frac{\Theta}{a} \, \dot h'_R\  ~~~, 
}
\ee
where
\be\label{Theta}
M^2_{pl}\Theta=4(F''\dot\phi^2+2HF'\dot\phi).
\ee
The unspecified function F is constrained ultimately by its direct relationship 
with the observed baryon to entropy asymmetry
\be 
\frac{n}{s} \sim 6.5 \pm 0.4  \times 10^{-10}
\ee 
By evaluating the the Green's functions of the gravitons eq (\ref{hphc-e.o.m}) in 
conformal coordinates
\be
           \eta = {1/H a}  = {e^{-Ht}/H} \ .
\ee
become
\be\label{Gequation}
  \left[{d^2\over d\eta^{2}} - 2 ({1\over \eta}+k\Theta) {d\over d\eta} 
+ k^2 \right] G_k(\eta,\eta')  = i  {(H\eta)^{2}\over M_{Pl}^{2}} \delta(\eta - \eta ').
\ee

For $\Theta = 0$, the solution of this equation is
\be\label{Gvalue}
   G_{k0}(\eta, \eta') = \left\{\begin{array}{cc}
             (H^2/2k^3 M_{Pl}^{2}) h_{L}^{+}(k,\eta) h_R^{-}(-k,\eta') 
                   &   \quad \eta < \eta' \cr
               (H^2/2k^3 M_{Pl}^{2}) h_{L}^{-}(k,\eta) h_R^{+}(-k,\eta') 
                   &   \quad \eta' < \eta\ , 
\end{array}\right. 
\ee
where $h_L^{-}$ and $h_R^{+}$ are the negative and positive frequency solution 
to the gravity wave equation (\ref{hphc-e.o.m}).  For $\Theta = 0$, these solutions 
are the same as for $h_L$.   After some more algebra we find the Greens function.
\be\label{myGk}
G_k=e^{-k\Theta\eta} G_{k0} e^{+k\Theta \eta'} 
\ee
By contracting the Greens function we obtain the quantum expectation value 
of ${\cal R}\wedge {\cal R}$   The answer is:
\be 
\label{RRdualval}
  \int d^{3}x <{{\cal R} \Tilde {\cal R}}(x)> =  {16\over a}\, \int \, 
  {d^3 k\over (2\pi)^3}\ {H^2\over 2 k^3 M_{Pl}^2} 
  (k\eta)^2 \cdot k^4 \Theta
\ee
As is well-known \cite{AGW},  the lepton number current, and also the 
total fermion number current, has a gravitational anomaly.  Explicitly,
\be
\label{Jlepton}
\partial_\mu J^\mu_\ell  =  {1\over 16\pi^2}   {\cal R} \Tilde {\cal R}
\ee
where
\be
J^\mu_\ell =    \bar \ell_i\gamma^\mu \ell_i + \bar \nu_i \gamma^\mu \nu_i \  ,\ 
{\cal R} \Tilde {\cal R} = \half \epsilon^{\alpha\beta\gamma\delta}
 {\cal R}_{\alpha\beta \rho\sigma} {\cal R}_{\gamma\delta}{}^{\rho\sigma} \ .
\ee
Inserting (\ref{RRdualval}) into (\ref{Jlepton}) and integrating over the time 
period of inflation, we find for the net lepton number density
\be
\label{netlept}
n = \int^{H^{-1}}_0 d\eta\ \int \, {d^3 k\over (2\pi)^3}\ 
{1\over 16\pi^2}\, {8 H^2  k^3 \eta^2  \Theta \over M_{Pl}^2}  \ .
\ee
The integral over $k$ runs over all of momentum space, up to the scale 
$\mu$ at which our effective Lagrangian description breaks down.  The 
dominant effect comes not from the usual modes outside the horizon at 
the end of inflation (super-horizon modes), $k/H < 1$, but rather from 
very short distances compared to these scales.  The integral over $\eta$ is 
dominated at large values of $\eta$, early times.  The integral represents 
a compromise between two effects of inflation, first, to blow up distances 
and thus carry us to smaller physical momenta and, second, to dilute the 
generated lepton number through expansion.   It is now clear that the dominant 
contribution to the right-hand side comes from $k\eta >> 1$, as we had 
anticipated.  Performing the integrals, we find
\be
\label{nval}
n  =  {1\over 72\pi^4} \left({ H\over 
M_{Pl}}\right)^2 \Theta H^3  \left({\mu\over H} \right)^6 \ .
\ee
We might interpret this result physically in the following way.  The factor
$(H/M_Pl)^2$ is the usual magnitude of the gravity wave power spectrum.
The factor $\Theta$ gives the magnitude of effective CP violation and is 
governed by the dynamics of the theory at hand. The factor $H^3$ is the 
inverse horizon size at inflation; this gives the density $n$ appropriate units.  
Finally, the factor $({\mu/H})^6$ gives the enhancement over one's first guess 
due to our use of strongly quantum,  short distance fluctuations to generate 
$<{\cal R} \Tilde {\cal R}>$, rather than the super-horizon modes which effectively behave 
classically.

The crucial assumption in the above leptogenesis mechanism was the origin
and the specific form of the function $F(\phi)$. While the authors argued that 
such a function can be  generic in the low energy effective action of string theory, 
it was not specified.  From another perspective one can ask a related question.  
Is the axion field in the low energy 4D effective action of string theory constrained 
to have couplings of the above form and if so what are the constraints on its 
dynamics?  We shall discover that not only is such a term quite universal, but 
the the axion is required to be dynamical due to a very peculiar set of superspace 
constraints, namely the 'beta-function-favored constraints' $\beta$FCC\cite{Gates}.  
\section{Level Crossing from Inflationary Gravity Waves}

   The net lepton number that was calculated in \cite{peskin} relied on the Atyiah-Singer (AS) index theorem.  The index theorem relies on Manifolds with definite topology.  In the inflationary slicing  of De Sitter space, the topology is indefinite so applying the Atyiah-Singer index theorem  should be handled with care.  One way of addressing the validity of the lepton number calculation is to perform a level crossing analysis of fermions in the background of the gravitational waves. Our goal in this section is to show how the gravity waves may give
rise to anomalous production of fermions via level crossing. In particular, if we have
gravity waves which evolve
in such a way that $R\tilde R$ has a non-vanishing integral, as is the
case in our model,
the fermion energy levels can cross from a negative to positive energy
spectrum. This would explicitly establish the relation between fermion
number generated via the presence of gravity waves rather than the usual topological arguments. In fact this is another way of
obtaining the gravitational ABJ anomaly result.

The idea is that instead of performing the (gravity) triangle loop
integral, one may focus on the Dirac equation in the background of the
gravity waves and directly find the energy levels and
how they evolve in time.

We begin by the action in which the (chiral) spin 1/2 fermion is
covariantly coupled to gravity:
\be\label{L-fermion}
L_{f}=({\det\ e})(\bar{\Psi_L}i \gamma^{\mu}\nabla_{\mu}\Psi_L)
\ee
where
\be
\nabla_{\mu}= \partial_{\mu} +\frac{1}{4} \omega_\mu^{ab}\gamma_{ab}
\ee
and $\omega_\mu^{ab}$ is the spin-connection. We will use Greek indices
for the space-time and
Latin indices for the tangent space and  as usual vierbeins $e^a_\mu$
relate these two and
\be
\eta_{ab}e^a_\mu e^b_\nu=g_{\mu\nu}
\ee
The spin-connection can be solved in terms of vierbeins using the identity
\be
 \label{spinc} D_\mu e^a_\nu=\partial_\mu e^a_\nu -\Gamma^\rho_{\mu\nu} e^a_\rho
+\omega_\mu^{ab} e_{\nu b}\equiv 0
\ee
where $\Gamma^\rho_{\mu\nu}$ is the Christoffel symbol.

Noting that $\gamma_\mu=e^a_\mu\gamma_a$ and the fact that
\be
\gamma^a\gamma^{bc}=\frac{i}{2}(\eta^{ab}\gamma^c-\eta^{ac}\gamma^b)-
\epsilon^{abcd}\gamma^5\gamma^d
\ee
we obtain
\be
L_{f}=({\det\ e})(\bar{\Psi_L}i\gamma^{\mu}\partial_{\mu}\Psi_L+ {\cal
L}_{int})
 \ee
with
\be\label{Lint}
{\cal L}_{int}=-\frac{1}{4}\bar\Psi_L (\omega_a\gamma^a +i
\tilde\omega_a \gamma^5\gamma^a)\Psi_L+c.c.
\ee
where
\be\label{omegaa}
\omega_a=e^\mu_{a}\omega_\mu^{ab}\ ,\
\ \ \tilde\omega_d=\epsilon^{abcd}e^\mu_{a}\omega_\mu^{bc} .
\ee
For the chiral fermions $\gamma^5\Psi_L=\Psi_L$ and hence
\be\label{Lintleftright}
{\cal L}_{int}=\frac{1}{4}\left[
\bar\Psi_L (\omega_a+i\tilde\omega_a) \gamma^a \Psi_L \right]+c.c.
\ee

For our case the metric can be parameterized as \be\label{metric}
ds^2=a^2(\eta)\left[-d\eta^2+ (\delta_{ij}+h_{ij})dx^idx^j \right]
\ee where $h_{ij}$ is a symmetric, traceless  and divergence-free
tensor parameterizing the gravity waves. For the above metric, the
spin connection can be written as \be\label{spin-connection}
\omega_\mu^{ab}=\left(\omega_\mu^{ab}\right)_0+\left(\omega_\mu^{ab}\right)_h
\ee where $\left(\omega_\mu^{ab}\right)_0$ is the contribution to
spin connection from the FRW metric (i.e. when $h_{ij}=0$) and
$\left(\omega_\mu^{ab}\right)_h$ is the contribution from the
gravity waves. Here we will only work in the first order in $h$'s,
and hence $\left(\omega_\mu^{ab}\right)_h$ is proportional to the
first order derivatives of $h$'s.

In order to study the fermion level crossing, we study the Dirac equation:
\be\label{eqm}
\gamma^\mu \nabla_{\mu}\psi=\gamma^\mu (\partial_\mu
+\frac{1}{4}\omega_\mu^{ab}\gamma^{ab})\psi=0
\ee

Next we note that the Dirac equation for the massless fermions is
conformally invariant, i.e. for any given  solution to the
background $g_{\mu\nu}$, $(\psi, g_{\mu\nu})$,
$\lambda^{\frac{d-1}{2}}\psi$ is also
a solution to the Dirac equation on the background with metric
$\lambda^{-2} g_{\mu\nu}$, e.g. see \cite{gibbons}.\footnote{ This
statement is not valid in the region (or locus) where $\partial_\mu
\ln\lambda$ is vanishing.}  Hence we can simply focus on
the gravity
waves in the flat background, and then multiply the solution to the
corresponding Dirac equation by $a^{3/2}$.

Let us now specifically solve the eigenvalue problem from the
above Dirac equation. %
\be \label{eigen1}%
(\gamma^{i}k_i +
\omega_{\mu}^{ab}\gamma^\mu [\gamma_{a},\gamma_{b}])\Psi_{D} = \gamma^0 E \Psi_{D}%
 \ee%
Since we are studying the phenomenon of level crossing we are in
the adiabatic regime of the quantum mechanics.  If the eigenvalues
adiabatically cross one another, then fermion creation is
established and we have level crossing across the Dirac sea. We
will aim to find the eigenvalues of the dirac equation in the
gravity wave background during the complete inflationary epoch.
From equation (17) in our paper, we have the solution for the
background gravitational wave    :

 \be\label{gdef}
     h_L = e^{ikz} \cdot (-ik\eta)
                        e^{k\Theta\eta}  g(\eta)
\ee
where \be\label{findg}
    g(\eta) =   \exp[ ik(1-\Theta^2)^{1/2} \eta (1 + \alpha(\eta))] \ ,
\ee
where $\alpha(\eta) \sim \log \eta/\eta$.
And the right handed component of the gravity wave can be obtained by changing the sign of $\Theta$ in $h_{L}$.

 We can solve explicitly for the non vanishing
components of the spin connection from eq (\ref{spinc}).  For our
specific situation we will consider the non-vanishing time-like
component of the spin connections.  In other words we will choose
a gauge where both the fermion and gravity plane waves are moving
in the $z$-direction, without loss of generality.   We find for
the spin connection after much of algebra,%
\be \omega_{0}^{12}=-i( h'_{L} -h'_{R})/2
\ee%
where $'$ denotes a derivative with respect to conformal time.
Plugging (\ref{gdef}) into (\ref{eigen1}) we get
\be %
\gamma^{3}\left[k + k\gamma^{5}\left(-k\Theta \eta \cosh(k\Theta
\eta) + (1-ik\eta)\sinh (k\Theta\eta)\right )\right]\Psi_{D}=\gamma_{0}E\Psi_{D}%
 \ee%
in our mechanism to get successful leptogenesis $\Theta \ll 1$,
and also we discussed in our paper the main contribution comes
from the region $k\eta\gg 1$ while $k\eta\Theta\ll 1$. So in
the
leading order the above eigenvalue equation takes the form%
\be%
\gamma^{3}k\left[1 - i \gamma^{5}(k\Theta \eta) k\eta\right]\Psi_{D}=\gamma_{0}E\Psi_{D}%
 \ee
and hence lepton number production when%
\be \label{cond}%
(k\Theta\eta)k\eta \rightarrow 1
\ee%
which is of course possible within the ranges of $k\eta \sim
10^3-10^4$ and $\Theta\sim 10^{-8}-10^{-9}$ that is considered in
\cite{peskin}, in accord with the recent cosmological data. We stress that this condition is exactly the regime considered in \cite{peskin} and is
consistent with lepton number production  via computing the
quantum expectation value of $\langle R\wedge R\rangle$ (provided
that $\Theta \ll 1$).  The authors of \cite{peskin} found the Greens functions
for the modes at short distances, $k\eta \gg 1$ which dominated
lepton number. It is indeed encouraging that the level crossing
analysis yields the same condition for lepton number production.
\footnote{ In the Ref. \cite{gibbons}, the computations has been
done for FRW background in the ``global'' coordinates, the
coordinates in which the metric is conformal to Einstein static
Universe, $R\times S^3$.}

\section{A Supergravity Realization}

~~~~ We begin our analysis from the compactification of the heterotic string to 
its 4D, ${\cal N} \,=\,1$ supergravity limit.  For concreteness we consider the 
compactification to be on a Calabi-Yau 3-fold or a deformed conifold.  In order 
to discuss cosmology in this context it is important to stabilize all moduli since 
they can lead to disasterous relic overproduction in the early universe since
this will be inconsistent with the bounds of nucleosynthesis.  Some attempts 
in the heterotic compactification to 4D has been made with the inclusion of 
fractional fluxes on a 3-cycle embedded the $CY_{3}$-fold \cite{Gukov}.  We 
will assume that all moduli except the axion are stabilized.

Our starting point is the 10D Heterotic string action in Einstein frame \cite{berg}.
\be \eqalign{
S=  \int d^{10}x \sqrt{g_{10}} \Big( \, &{\cal R} -{1\over 2} \partial_{A} \phi 
\partial^{A} \phi -{1\over12} e^{-\phi} \, H_{ABC}  ~
-~ {1\over4}e^{{-\phi\over2}}tr(F_{AB}F^{AB})   ~ \Big) }   \ee
where 
\be 
H_{3} = dB_{2} - {1\over4}\big(\Omega_{3}(A) - \alpha'\Omega_{3}(\omega) 
\big) 
\ee
where $\Omega_{3}(A)$ and $\Omega_{3}(\omega)$ are the gauge and 
gravitational Chern-Simons three-forms respectively.   
\be 
\Omega_{3}(A)= Tr \big(dA\wedge A + {2\over 3}A \wedge A \wedge 
A \big) 
\ee
We now dimensionally reduce the 10D action to 4D, ${\cal N} \,=\, 1$ supergravity 
coupled to the $SO(32)$ or $E_{8}\times E_{8}$ gauge sectors by choosing a 
four-dimensional Einstein frame metric , $g^{S}_{MN}=g^{E}_{MN}e^{{\phi\over
2}}$.
\be 
ds^{2}_{10}=e^{-6\sigma}ds_{4}^{2} + e^{2\sigma}g_{mn}dy^{m}dy^{n} 
\ee
where $g_{mn}$ is a fixed metric if the internal dimensions normalized to have volume $4\alpha'^{3}$
\be 
S_{4D}={1\over 2\kappa_{4}^{2}}\int d^{4}x\sqrt{-g}\big[ \, {\cal R} \,-\, {2\partial_{
\mu}S^{*} \partial^{\mu}S \over(S+S^{*})^{2}} -{1\over2}G^{i\bar{j}}G^{k\bar{l}}
\partial_{\mu} T_{i\bar{l}}\partial^{\mu}T_{\bar{j}k} \big] 
\ee
The field $T_{i\bar l}$ are associated with moduli in the internal dimensions.  For our 
purposes we are assuming moduli stabilization so these fields will not be relevant.  For 
a further discussion of this issue in specific moduli stabilization scheme we refer the 
reader to \cite{Gukov}

It is now useful to relate the string scale and coupling to the four dimensional moduli.  
The four dimensional gauge coupling is 
\be 
g^{2}_{YM} = e^{\psi} .
\ee
where the four dimensional dilaton $\psi$ is related to the ten-dimensional dilaton 
and volume modulus via
\be 
\psi={\phi\over2} - 6\sigma 
\ee
Also related to the ten dimensional dilaton is the volume scalar $\rho$.
\be 
\rho ={\phi\over2} + 2\sigma 
\ee
The fields $\psi, \rho$ are related to the scalar components of the two \footnote{
Due to our choice of metric ansatz the gravitational coupling is $\kappa_{4}^{2} 
= {\alpha'\over 4}$.  The physical  \newline $~~~~~~$
gravitational coupling differs from this by a 
constant rescaling.} $\cal{N}\rm=1$ chiral superfield $S$
\be 
S=e^{-\psi} + ia 
\ee

where $a$ is the model independent axion field which arise from the spacetime and internal components of 
$B_{AB}$ respectively.  Specifically, 
\be 
(*da)_{\mu\nu\rho} =e^{-2\psi}H_{\mu\nu\rho} 
\ee

The heterotic string compactified to 4 dimensions can exhibit some moduli 
stabilization due to fluxes, hence we refer the reader to the work of Gukov et. al. 
\cite{Gukov}. In what follows, we focus our attention on the axionic sector of the the 4D heterotic string.  
The bosonic low energy effective action will take the form
\be 
S_{4d} = S_{gravity} \\ + \\ S_{axion} \\ + \\ S_{CS} 
\ee
\be 
S_{4d}  = {2\over\alpha'}\int d^{4}x\sqrt{-g}\big( \, {\cal R}_{4} + {1\over2} \partial_{\mu}
a \, \partial^{\mu}a + V(a) \big) 
\ee
where $V(a)$ is the potential for the axion which will be responsible for inflation.  We 
shall discuss the form of this potential in the following section.  Finally, there will be an 
important contribution from the Green-Schwartz mechanism 
\cite{GS}
\be 
\label{f} \int d^{4}x F(a){\cal R}\wedge {\cal R} 
\ee
where
\be 
F(a) = a\cal{V}\rm M_{4pl}\alpha' 
\ee
where $\cal{V}$ is a volume factor measured in string units and is determined on the 
dimensionality of the compactification.  Take note that eq (\ref{f}) is precisely the term 
which is needed for our inflationary leptogenesis to occur.  Most importantly this 
interaction and its coupling is universal. 

\subsection{Natural Heterotic Inflation}

~~~~ In the previous section we demonstrated that a 4D effective description of the 
heterotic string can have a dynamic axion.  In the presence of potential which 
satisfies the flatness conditions for inflation, the axion can act as the inflaton field.  
In general, axions acquire oscillatory potentials reflecting the periodicity of the 
$\Theta$ vacua.  These sorts of potentials were first discussed as inflationary 
potentials by Freese et. al and extended to Supergravity by Kawasaki et. al under 
the name "Natural Inflation".   These potentials are constrained from the superpotential 
in ${\cal N}\,=1$ supergravity and are naturally flat due to the shift symmetry in the 
chiral superfield.  In particular, the F-term potential in Planck units is 
\be 
V_{F}=e^{K} \big(\sum_{i,j} K^{i,j}D_{i}WD_{j}W-3|W|^{2}\big), 
\ee where $i,j$ 
runs over all moduli fields, $K$ is the Kahler potential for $T$ and $S$, $K^{i\bar{j}}$ 
is the inverse of $\partial_{i}\partial_{\bar{j}}K.$ These authors \cite{Yamaguchi,
Adams} found flat potentials which could generate inflation.  Recently, Blanco-Pillado et. al.
have demonstrated a general class of axionic potentials which generate inflation as 
an extension to the KKLT mechanism\cite{Kachru,Burgess}.  It will be interesting to 
realize a similar scenario in our heterotic example. 

It is not necessary that the axion be identified with the inflaton field, however.  
Inflation could be driven by another field and the axion, through its equation of motion 
in this background can track the slow rolling of the inflaton field.
\be 
\ddot{a} + 3H_{inf}\dot{a} = -V'(a) 
\ee
where $H_{inf}$ is the Hubble parameter associated with the inflaton field rather than 
the axion.  Since our mechanism only requires the axion to be coherent over the 
Hubble scale during the course of inflation, this amounts to it having a small slow roll
paramenter, $\epsilon ={1\over2}({\dot{a}\over H_{inf}M_{pl}})^{2}<<1$.  This equation 
arises from the equation of motion of the axion during inflationary expansion.  In other 
words the axion may have a flat potential to guarantee its slow rolling but not necessarily 
have enough energy density to yield enough inflation.  The important point for our analysis  
is that we have inflation and its associated scale since our calculation involves 
constraining the Hubble constant during inflation from recent CMB measurements.
 
\section{An Observational Constraint on the String Scale}

~~~~ We are now ready to determine the string length scale and coupling by relating 
it to observed baryon to entropy ratio.  The crucial point is that any four dimensional 
compactification of string theory is endowed with the model independent axion which 
couples to ${\cal R}\wedge {\cal R}$ via the Green-Schwarz mechanism.  It is through this fact that 
we are able to uniquely observationally constrain fundamental string parameters.  Recall 
that the baryon to entropy ratio is fully determined by the gravitational power spectrum, 
the CP-violation term which enhances the amplitude of the gravitational waves and the 
momentum dependence of the gravitational waves.  The CP-violation term $\Theta$ is 
fully controlled by the dynamics of the fundamental theory.  We shall obtain a result 
which constrains the ratio between $M_{pl}$  and $M_{10}$, the fundamental and four dimensional Planck scale, respectively, through the observed baryon to entropy ratio.   

In the previous section we motivated a heterotic extension of inflationary leptogenesis
via. gravitational waves which experience birefringence during inflation.  The baryon
to entropy ratio was calculated and determined in terms of three terms, each signifying distinct physics. 
\be 
\label{ns} {n\over s} = 10^{-6} \Theta ({H\over M_{pl}})^{7\over2} ({\mu\over H})^{6} 
\ee
where ${H\over M_{pl}}$ is constrained by WMAP to have an upper bound of
${H\over M_{pl}} \sim 10^{-4}$.  Recall that we can express $F(a)$ from the equation 
of motion of the gravitational waves in terms of $\Theta$.

\be
\label{Theta}
M^2_{pl}\Theta=4(F''\dot\phi^2+2HF'\dot\phi)\cal{V}\rm.
\ee  
Because the axion is slowly rolling during inflation and is linearly coupled to the 
Chern-Simons form, the first term vanishes and we obtain.
\be 
\Theta = 4F'H\dot{a}{1\over M_{pl}^{2}}\cal{V}\rm 
\ee
This term is completely determined by string theory since it involves the model independent 
axion.  Therefore, the string scale and coupling will determine the observed baryon to 
entropy ratio.  We can simplify Theta by using the relation between the slow roll parameter 
and the velocity of the axion field.  After a little algebra we get,
\be \label{vol}
\Theta = ({H\over M_{pl}})^{2} \cal{V}\rm \sqrt{\epsilon} .
\ee
Notice that $\cal{V}$ the volume factor can lead to an enhancement in the Greens function 
of the gravity waves.  This could be important for future CMB observations of gravity waves, 
since one could obtain an enhancement in the B-mode polarization in the gravity wave 
power spectrum.  We are currently investigating this possibility by extending the work of Lue 
et. al \cite{Kamion,shahin}   Specifically, $\cal{V}\rm= {M_{pl}\over M_{10}}(Vol M^{6}_{10}) $ 
and it is useful to recall that
\be 
M_{pl}^{2}=M^{8}_{10} Vol 
\ee 
and
\be 
M_{10}^{8}=\alpha'^{-4}g_{s}^{2} 
\ee
where $Vol$ is the volume of the Calabi-Yau 3-fold.  Hence,
\be 
\cal{V}\rm = ({M_{pl}\over M_{10}})^{3} 
\ee
with the above definitions eq (\ref{ns}) become
\be 
{n\over s}= 10^{-6}({H\over M_{pl}})^{1/2}({\mu\over H})^{6} 
({M_{pl}\over M_{10}})^{3}  
\ee
We can set the cutoff scale of the Chern-Simons interaction to be the string scale, $\mu=M_{
10}$ and and demanding from observations that
\be 
{n\over s} \sim 10^{-10} 
\ee
we get finally
\be 
{M_{10}\over M_{Pl}} \sim 10^{-2} 
\ee
which sets the string scale to be 
\be 
{g_{s}^{1/4}\over M_{pl}l_{s}} \sim 10^{-2} 
\ee
We immediately get a lower bound for the string scale to be aroun $10^{17} GeV$ provided 
that the string coupling is order unity.  Clearly as we weaken the string coupling the 
string  scale goes down.      

\section{Conclusion}

~~~~ In this work we have provided a stringy embedding of the inflationary leptogenesis 
mechanism of \cite{peskin}.  This was possible due to the universality of the coupling 
of the model independent axion to the Chern-Simons form in four dimensional 
compactifications of string theory.    In the original work of \cite{peskin} the baryon to 
entropy ratio required an unspecified enhancement to accommodate the observed 
value of ${n\over s} \sim 10^{-10}$.  In our realization this enhancement arose naturally 
due to the volume factor in eq (\ref{vol}).  We chose to explicitly study the heterotic string 
theory for concreteness and found that for reasonable values of the string coupling and 
length scale the observed baryon to entropy ratio can be generated during a period of 
inflation.  

We have presented a fairly conservative compactification.  If the volume of compactification 
is larger than the string scale, such as in warped compactifications, then the enhancement 
of the baryon asymmetry will be improved.  It will be interesting to further investigate this 
mechanism in the context of brane world scenarios and stringy inflationary mechanisms 
driven by axions which were already studied by Kallosh et al and Blanco-Pillado et al.  

In this leptogenesis mechanism the UV modes of gravitational waves coupled to volume enhancement factor $\cal{V}\rm$ are responsible for the enhancement of the lepton number density.      In future CMB experiments such as PLANK the "smoking gun" of inflation, the tensor to scalar ratio $r={T\over S}$ will be constrained for models with high scale inflation\cite{Lyth}.  Likewise the tensor power-spectrum can be enhanced by the volume factor from string theory.  In is intriguing to relate this mechanism directly to the the IR (superhorizon) gravitational power specturm associated with the CMB.  We expect to report on this  connection in a future paper \cite{jerome}   

\section{Acknowledgements}
~~~~ It is our pleasure to thank especially Mohammad Sheikh-Jabbari (Shahin) for many 
discussions and checking a draft of this paper.  We also give thanks to Robert 
Brandenberger and Michael Peskin for reading a draft of our paper and discussions. 
\\[.5in]
$~~~~~~$ {\it ``Hidden deep in the heart of strange new elements are
secrets beyond \newline $~~~~~~\,~$
human understanding, new powers, new dimensions --
world within  \newline $~~~~~~\,~$ worlds unknown."}

 $~~\,$ Quote from Outer Limit Episode: ``The Production and
 Decay of   \newline  $~~~~~~\,~\,$ Strange Particles''

\end{document}